# White Paper

on the case for

# LANDED MERCURY SCIENCE

**Paul Byrne** NC STATE UNIVERSITY | paul.byrne@ncsu.edu

**David Blewett** APPLIED PHYSICS LABORATORY

**Nancy Chabot** APPLIED PHYSICS LABORATORY

**Steven Hauck, II** CASE WESTERN RESERVE UNIVERSITY

**Erwan Mazarico** NASA GODDARD SPACE FLIGHT CENTER

**Kathleen Vander Kaaden** JACOBS/NASA JSC

**Ronald Vervack** APPLIED PHYSICS LABORATORY

**ENDORSERS** ROSALIND ARMYTAGE | BRENDAN ANZURES | W. BRUCE BANERDT | MARIA BANKS | JOHANNES BENKHOFF | SEBASTIEN BESSE | NICOLAS BOTT | JOSHUA CAHILL | CRISTIAN CARLI | CLARK CHAPMAN | EDWARD CLOUTIS | GABRIELE CREMONESE | BRETT DENEVI | ARIEL DEUTSCH | CHUANFEI DONG | ALAIN DORESSOUNDIRAM | NICHOLAS DYGERT | DENTON EBEL | CAROLYN ERNST | CALEB FASSETT | LORENZA GIACOMINI | JEFFREY GILLIS-DAVIS | SANDER GOOSSENS | CESARE GRAVA | BENJAMIN GREENHAGEN | JENNIFER GRIER | JAMES HEAD | ANDREA HERRERO-GIL | DANIEL HEYNER | GEORGE HO | JOSE HURTADO, JR. | HAUKE HUSSMANN | LUCIANO IESS | SUZANNE IMBER | CATHERINE JOHNSON | MATTHEW IZAWA | NOAM IZENBERG | PETER JAMES | STEPHEN KANE | LAURA KERBER | MALLORY KINCZYK | SCOTT KING | RACHEL KLIMA | CHRISTIAN KLIMCZAK | JURRIEN KNIBBE | DAVID LAWRENCE | FRANK LEMOINE | ALICE LUCCHETTI | FRANCIS McCUBBIN | RALPH McNUTT, Jr. | MOHIT MELWANI-DASWANI | JAMIE MOLARO | CATHERINE NEISH | LARRY NITTLER | JULIE NEKOLA NOVÁKOVÁ | JÜRGEN OBERST | CSILLA ORGEL | LILLIAN OSTRACH | SEBASTIANO PADOVAN | MAURIZIO PAJOLA | MARK PANNING | STEPHEN PARMAN | PATRICK PEPLOWSKI | PARVATHY PREM | SOUMYA RAY | JOE RENAUD | KURT RETHERFORD | EDGARD RIVERA-VALENTÍN | STUART ROBBINS | JAMES ROBERTS | THOMAS RUEDAS | RICHARD SCHMUDE, JR. | NORBERT SCHÖRGHOFER | MORGAN SHUSTERMAN | MATTHEW SIEGLER | MARTIN SLADE | JAMES SLAVIN | RYAN SMITH | ALEXANDER STARK | BECK STRAUSS | HANNAH SUSORNEY | MICHELLE THOMPSON | ARYA UDRY | INDHU VARATHARAJAN | FAITH VILAS | AUDREY VORBURGER | JENNIFER WHITTEN | ZOE WILBUR | DAVID WILLIAMS | REKA WINSLOW | ZHIYONG XIAO

**Key Findings**
- **We advocate for establishing key scientific priorities for the future of Mercury exploration—including the development of specific science goals for a landed mission**
- **We support the Mercury science community in fostering closer collaboration with ongoing and planned exoplanet investigations**
- **The continued exploration of Mercury should be conceived as a multi-mission, multi-generational effort, and that the landed exploration of Mercury be a high scientific priority in the coming decade**

## 1. Current and Planned Mercury Exploration

The arrival at Mercury in 2011 of NASA's MESSENGER mission heralded a new age of exploration for this enigmatic planet (**Figure 1**). The MESSENGER (MErcury Surface, Space ENvironment, GEochemistry, and Ranging) spacecraft (Solomon *et al.*, 2008) operated at Mercury for over four Earth years, acquiring global observations of the planet's surface and measurements of the interior, exosphere, and magnetosphere. Thanks to MESSENGER, we now know Mercury to be a world once extraordinarily geologically active but that has some surface processes that persist even today. It is also a planet with a composition and interior structure unlike that of the other terrestrial bodies in the Solar System, and which hosts complex interactions between an intrinsic magnetic field and a dynamic heliospheric environment. Our understanding of Mercury will be enhanced further by the 2025 arrival of the joint ESA–JAXA BepiColombo mission (Benkhoff *et al.*, 2010); with two discrete spacecraft, BepiColombo will characterize in greater detail the planet's surface, interior, and the magnetosphere–interplanetary solar wind interaction.

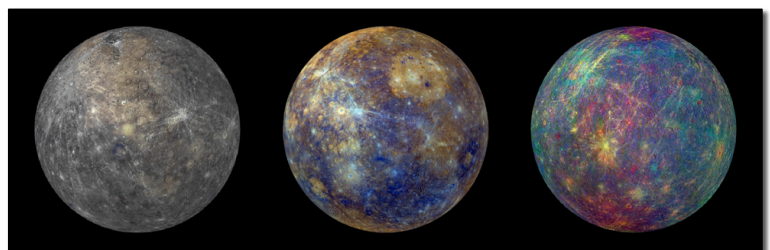

*Figure 1. The MESSENGER spacecraft returned unprecedented, global views of Mercury including, from left to right, color (1000, 750, and 430 nm in red, green, and blue), enhanced color, and multispectral data.*

Yet there is a limit to the scientific return of an orbital mission: an orbiter cannot directly sample surface materials, for example, nor is it able to delve into the interior in the way that a landed mission can. Indeed, the planetary science community has long adopted a stepwise strategy of exploration that starts with flybys before moving to orbiters, and then to landers, rovers, and, ultimately, sample return (NRC, 2011). Mercury was visited first by the NASA Mariner 10 spacecraft, which performed three flybys of the planet in the 1970s. With the successful completion of the MESSENGER mission, and the arrival in the coming decade of BepiColombo, our exploration of Mercury has accomplished the first two phases of this stepwise strategy. It stands to reason, then, that we should begin to consider the benefits of a landed mission at Mercury.

Here, we identify several key aspects of Mercury science that can be best addressed by such a mission. **Our goal here is not to advocate for a specific location on Mercury, but to demonstrate why such a lander in general would represent a natural next step in the exploration of this planet.** Detailed determination of Mercury's composition, evolution, and interaction with its space environment are crucial for addressing the planetary science community's priorities to understand the beginnings of planetary systems and how planets evolve through time (NRC, 2011). To leverage our growing knowledge—and increasing depth—of the other inner Solar System bodies, it is necessary to develop a comparably deep understanding of Mercury.

We must therefore plan for a steady stream of missions to the innermost planet over the coming decades, with each building upon its predecessor. With the necessarily long cruise time from Earth, comparable to destinations in the outer Solar System, and the limited number of spacecraft mission opportunities, **the time to consider landed exploration of Mercury is *now*.**

## 2. The Case for Landed Mercury Science

In this section, we discuss several major aspects of Mercury's character and evolution where substantial knowledge gaps exist, but where our current understanding could be dramatically improved with data acquired from the planet's surface. We do not offer specific recommendations for any particular landed mission architecture, but we note where appropriate potential types of instrumentation that could aid in addressing these gaps. **We emphasize that this discussion, though illustrative, is by no means exhaustive.**

## 2.1. Geochemistry: Placing Mercury in Geochemical Context with Other Rocky Worlds

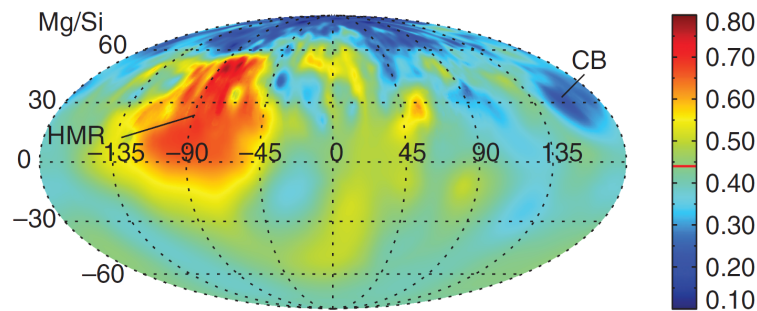

*Figure 2. Mg abundance on Mercury. Map is in Mollweide projection, centered at 0°N, °E. Red line in color scale is area-weighted global average of mapped data. HMR: high-Mg region; CB: Caloris basin. After Nittler et al. (2018).*

Geochemical observations obtained by the X-Ray Spectrometer (XRS) and Gamma-Ray and Neutron Spectrometer (GRNS) onboard the MESSENGER spacecraft revealed Mercury as a geochemical end-member among the rocky planets (e.g., Nittler et al., 2011; Peplowski et al., 2011). The high abundances of sulfur (>3 wt%) and low abundance of iron (<3 wt%) on the surface of Mercury indicate extremely low oxygen fugacity, such that Mercury is the most chemically reduced of the terrestrial planets (e.g., Nittler et al., 2011; Zolotov et al., 2011; McCubbin et al., 2017). In oxygen-starved systems, elements will deviate from the geochemical behavior that they exhibit at higher oxygen fugacities. In situ geochemical analyses would give new insight into these behaviors, allow for better interpretations regarding the thermochemical evolution of the planet, and provide substantial advances toward our understanding of planet formation.

Mercury is extremely diverse in terms of surface compositions (e.g., Peplowski et al., 2015a; Weider et al., 2015; Vander Kaaden et al., 2017) (**Figure 2**) and is also volatile-rich (e.g., Peplowski et al., 2011), an unexpected finding given the planet's heliocentric distance (e.g., Albarède, 2009; Peplowski et al., 2011; Peplowski et al., 2014; Peplowski et al., 2015b). Yet despite the insights provided by MESSENGER and those sure to come from BepiColombo, several outstanding compositional questions remain, including:

- **the nature, origin, and abundance of Mercury's low-reflectance material**;
- **the mineralogy of the planet's varied surface materials**; and
- **the composition of diffuse deposits interpreted to be pyroclastic in nature**.

Placing tighter constraints on the geochemical, mineralogical, and isotopic properties of the surface can be accomplished through in situ compositional and petrological measurements obtained from a lander mission equipped with geochemical and imaging instruments. Given Mercury's geochemical end-member characteristics, the results obtained from landed science would give us unprecedented information on planetary differentiation and formation processes in our Solar System—information that could also be used as a local analog for understanding extrasolar planets, and particularly those close to their host star. A fuller understanding of Mercury's geochemistry would also inform subsequent exploration efforts, especially the aspirational goal of sample return from the planet (Vander Kaaden et al., 2020), and could even help to identify samples from Mercury proposed to exist in the worldwide meteorite collection (e.g., Gladman and Coffey, 2009).

## 2.2. Interior Structure: Understanding Planetary Formation in the Solar System

With its high bulk density (Ash et al., 1971) and super-size metallic core (Smith et al., 2012) (**Figure 3**), Mercury occupies a unique place among terrestrial planetary bodies and is key to understanding planet formation and evolution. The origin of Mercury is indeed still unclear, particularly its high metal-to-silicate ratio. Improved geophysical constraints in addition to new in situ geochemical data are needed to refine or discard existing "chaotic" and "orderly" formation models (Ebel and Stewart, 2018).

**Crucial geophysical data could be effectively acquired by a landed mission.** For example, a lander equipped with a seismometer would provide:

- **a determination of the interior structure with high fidelity**;

- **important constraints on density, temperature, and composition at depth**; and
- **the present-day level of seismicity at Mercury**.

The degree of seismic activity on Mercury is unknown; however, the planet undergoes thermal cycling (Williams *et al.*, 2011), flexing from solar tides (e.g., Padovan *et al.*, 2014), and may even still be contracting (Banks *et al.*, 2015)—and these crustal processes could be assessed with a seismic investigation. The present impact flux at Mercury could also be characterized, placing vital bounds on the impact history of the inner Solar System (e.g., Le Feuvre and Wieczorek, 2011). Although multiple stations would be preferable, the NASA Discovery-class InSight mission (Banerdt *et al.*, 2012), operating on Mars since November 2018, has demonstrated the capability of single-seismometer experiments for interior studies. And a single seismic station might perform better on a world with as shallow a core as Mercury.

A landed mission would also offer an opportunity for high-accuracy geodesy, as direct-to-Earth radio tracking would help improve the orientation dynamics—in particular, the longitudinal librations and the nutation of the spin axis (especially for a landing site at low latitudes), which are sensitive to the size and shape of the core (Dehant *et al.*, 2011). In addition to a seismometer and radio transponder, which could place further bounds on the size of the inner core (Genova *et al.*, 2019), other experiments could be advantageously included to make the lander a geophysical station. For example, a heat probe (as for the InSight mission) would provide crucial heat flux observations directly relevant to the core dynamo (Stanley *et al.*, 2005) as well as to topography compensation mechanisms (James *et al.*, 2015). A magnetometer would help characterize the electrical and conductivity structure of the crust and mantle (Johnson *et al.*, 2016; Zhang and Pommier, 2017). And the science return of a Mercury geophysical lander would be further enhanced if paired with GRAIL-like orbiters (Zuber *et al.*, 2013) or a GOCE-like gravity gradiometer (Drinkwater *et al.*, 2003; Griggs *et al.*, 2015) to finely map density and thickness variations in the lithosphere and crust. Indeed, an orbiting laser ranging system for use with a laser retroreflector on the lander would yield even more accurate geodetic data.

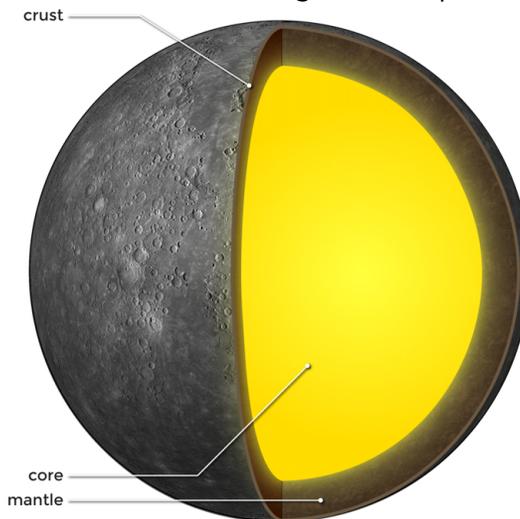

*Figure 3.* Schematic of the interior of Mercury. The core is more than 80% the radius of the entire planet (e.g., Margot et al., 2018).

## *2.3. Geological History: Exploring Mercury's Evolution since Formation*

Data returned by the MESSENGER mission have provided a global characterization of the history of the planet as recorded by its surface features (e.g., Denevi *et al.*, 2013; Marchi *et al.*, 2013; Byrne *et al.*, 2014). Mercury was an active planet early in its history, as evinced by its modest spatial density of large impact basins (Marchi *et al.*, 2013) followed by a rapid waning of volcanic activity (Byrne *et al.*, 2016), all of which are overprinted by tectonism associated with global contraction (Byrne *et al.*, 2014; Watters *et al.*, 2015).

However, as for all bodies beyond the Earth–Moon system, we still lack a complete understanding of the absolute ages of events, landforms, and deposits on the surface. **In situ geochronological measurements of surface materials with sufficient precision could place vital constraints on the absolute timing of events in Mercury's evolution**, as well as chronological and impact flux models for the entire Solar System.

As MESSENGER orbited closer to the surface near the end of its mission, crustal remanent magnetization was discovered (Johnson *et al.*, 2015; Hood *et al.*, 2016) (**Figure 4**). However, magnetization signals detected at orbital altitudes require magnetizations over considerable depth, and so an orbiter cannot provide the necessary insight into where such signals arise in the crust. Investigating remanent magnetization with a surface magnetometer on a landed mission would establish important links between:

- **surface geological processes and evolution**;
- **integrated igneous activity and depth**; and

- **the history of interior melt production and dynamo generation**.

Determining the carriers of magnetization (Strauss *et al.*, 2016) through geochemical and mineralogical assessment of surface materials (§2.1) is key to understanding crustal magnetization and its history. Such assessment, in concert with investigation of crustal structure with a seismic experiment (§2.2), would yield meaningful limits on estimates of the thickness of magnetization on Mercury—especially when paired with local magnetic field measurements. These local measurements would also aid complementary studies of electromagnetic fields in the crust and mantle to characterize internal structure (Anderson *et al.*, 2014; Johnson *et al.*, 2016) (§2.2), and interactions between the internal and external magnetic field (§2.4).

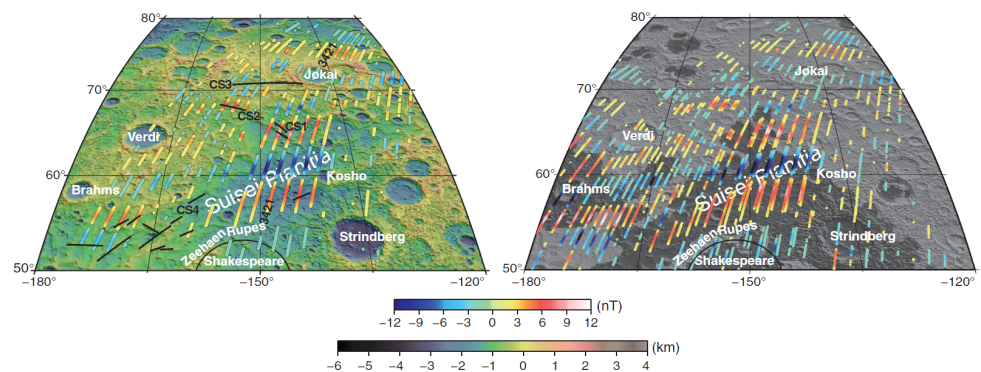

*Figure 4.* Remanent magnetic field detected in Mercury's crust. Signatures detected by MESSENGER over Suisei Planitia are shown. Crustal magnetization was detected both at altitudes of 25–60 km (left) as well as at lower altitudes of 14–40 km (right). After *Johnson* et al. *(2015)*.

## 2.4. Mercury Today: Investigating Active Planetary Processes

The MESSENGER mission showed us that Mercury experiences a number of active processes that could readily be investigated by instruments on a lander. For example, the surface is subjected to an especially harsh space-weathering environment (e.g., Domingue *et al.*, 2014). As these particle–surface interactions are an important source of the exosphere (e.g., Martinez *et al.*, 2017; Merkel *et al.*, 2018), and may contribute to macroscopic landscape modification in the formation of hollows (e.g., Blewett *et al.*, 2016), it is critical that we better understand the effects of solar-wind and magnetospheric charged particles (ions and electrons) and interplanetary dust particles (IDPs) on Mercury's surface materials. Although information on the charged particle environment surrounding the planet was obtained by MESSENGER, and will be substantially augmented by BepiColombo's dual-spacecraft measurements, **in situ measurements at the surface enable the direct study of particle–surface interactions**.

Measurements that are needed include, but are by no means limited to:
- **the incoming IDP flux at the surface**;
- **the flux of charged particles, both from the magnetosphere and solar wind as well as that released from the surface during sputtering and meteoroid impact vaporization events**; and
- **the neutral atoms and molecules present**.

The acquisition of these data could be accomplished with a combined ion and neutral mass spectrometer and a dust experiment. Together with in situ analysis of mineralogy and geochemistry (§2.1), these charged particle and IDP measurements would greatly further our understanding of the source and loss mechanisms behind the complex surface–exosphere–magnetosphere system, and of the processes involved in the initiation and growth of Mercury's distinctive hollows (***Figure 5***).

Mass spectrometers would also allow detection at the surface (and during descent) of exospheric density, a measurement crucial for determining both the high-mass-atoms composition of the exosphere and the release processes at work at the surface, and could also help characterize the absorption spectra of surface materials at Mercury conditions (Helbert *et al.*, 2013; Ferrari *et al.*, 2014). And in situ imaging of the surface could determine the physical properties of the regolith, including grain size, shape, and mechanical strength.

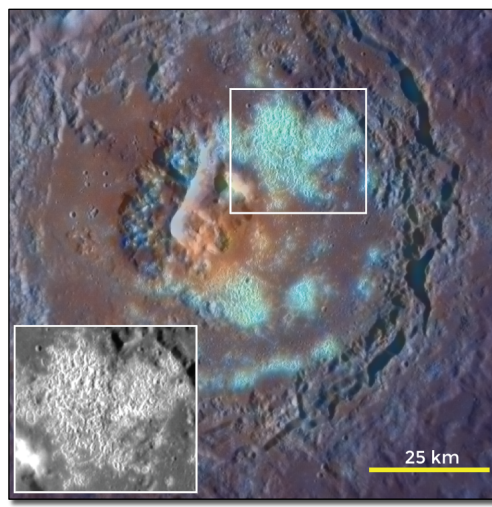

*Figure 5.* Enhanced-color view of hollows (blue) inside the 97-km-diameter Tyagaraja crater. Inset: hollows in monochrome. After *Blewett* et al. *(2011)*.

Moreover, large-scale investigations of the morphological structure and temporal dynamics of the exosphere and magnetosphere could be conducted from the surface. These measurements could be obtained using either an imaging spectrometer system to provide both spectral and spatial information, or by the use of an all-sky camera with narrowband filters. Such methods are routinely used to study the Earth's airglow, and could be similarly employed at Mercury. The siting of these instruments near the midnight equator would allow intense study of the tail structure, whereas a location near the poles would enable a study of the day–night transport. A fixed-surface location is desired because completely disentangling the spatial and temporal aspects from a rapidly moving spacecraft is difficult—another example of how a Mercury lander could build upon the science return of previous and planned orbiter missions.

### 2.5. Polar Volatiles: Understanding Inner Solar System Volatile Inventories and Origin

Earth-based radio telescopes provided the first tantalizing evidence for the presence of water ice at Mercury's polar regions (e.g., Slade *et al.*, 1992; Harmon and Slade, 1992; Butler *et al.*, 1993; Harmon *et al.*, 2011). Subsequently, multiple MESSENGER datasets provided strong evidence that Mercury's radar-bright materials are composed of water ice: the deposits are located in permanently shadowed regions (e.g., Deutsch *et al.*, 2016; Chabot *et al.*, 2018) with temperatures cold enough to sustain water ice (Paige *et al.*, 2013); neutron spectrometer results show elevated levels of H in Mercury's north polar region (Lawrence *et al.*, 2013); and reflectance measurements and images have revealed the surfaces of the polar deposits to have albedo properties distinct from Mercury's regolith (e.g., Neumann *et al.*, 2013; Chabot *et al.*, 2016). Together, these data point to extensive deposits of water ice and other volatile compounds in Mercury's polar regions (Deutsch *et al.*, 2020) (*Figure 6*).

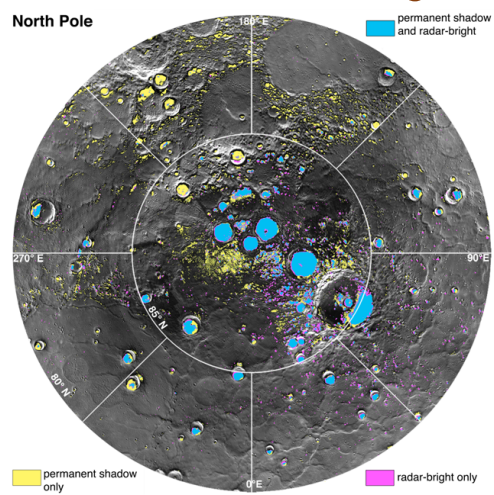

*Figure 6.* Mercury's polar deposits feature large expanses of exposed water ice, as well as other volatiles. North polar stereographic view. After Deutsch et al. (2016).

MESSENGER data confirmed that these **large deposits of volatiles are exposed directly on the surface**, providing a unique opportunity for landed science. In situ measurements are ideally suited to address the major open science questions about Mercury's polar deposits, including the origin of Mercury's polar volatiles, and whether the deposits represent an ancient, recent, or ongoing formation process; the nature of the volatiles trapped at Mercury's poles, and whether they include organic-rich materials delivered to the inner planets; and the processes that act in permanently shadowed regions, and if these processes produce or destroy water ice.

Addressing these questions has implications not only for Mercury but also for understanding the inventory of inner Solar System volatiles, including those on the Moon and the potential delivery of volatile species to early Earth and Mars. Landed measurements would provide fundamental new data not otherwise available to us, such as direct measurements of:

- **the composition of the volatile compounds within Mercury's polar deposits**;
- **the purity of the ice**; and
- **the physical properties of volatiles, including grain size, strength, thickness, and evidence for layering**.

Such measurements would address crucial, open science questions about Mercury's polar volatiles, which in turn would provide new insight into the volatile inventory and evolution of the inner Solar System worlds.

### 3. Outlook

Key to ensuring a firm footing for continued Mercury science is supporting the Mercury science community to organize and discuss the future priorities of the scientific exploration of Mercury. **We are encouraged by** recent NASA efforts to constitute the Mercury Exploration Assessment Group (MExAG) to enable the planetary science community to formulate and advocate exploration goals for the innermost planet. It is crucial that key scientific priorities now be established by the planetary science community, utilizing the MExAG, for the future of exploration of Mercury—including the development of detailed, specific Mercury science goals such as those listed in this White Paper. **By doing so, future mission concepts, such as those proposed at the openly competed Discovery and New Frontiers-class levels, would have strong scientific motivation backed by community-generated priorities.** These concepts will especially benefit from mission design studies performed in support of the 2023–2032 Decadal Survey, such as that by Ernst *et al.* (2020).

Our improved knowledge of Mercury now enables us to understand more fully the evolution of terrestrial planets in general, potentially including those in orbit about other stars. For example, it is possible that Mercury is an important model for extrasolar planets in high-C solar systems. Planets that are carbon rich are expected to have low oxygen fugacities, and may therefore feature sulfur-rich crusts and, if present, atmospheres. Mercury is also a useful analog for studying exoplanets with major iron mass fractions (e.g., Santerne *et al.*, 2018). **We therefore support the Mercury science community in fostering closer collaboration with ongoing and planned exoplanet investigations.**

Finally, the development and eventual dispatch to Mercury of a lander—whether targeting low latitudes, the poles, or a particular surface unit (e.g., Ernst *et al.*, 2020)—should not signify the end of exploration efforts for the planet. Indeed, following the decades-long established protocol of flyby, orbiter, and lander approach taken by NASA (NRC, 2011), it follows that an aspirational goal should be the collection from the surface and the delivery to Earth of a sample of Mercury (Vander Kaaden *et al.*, 2020). Such a sample would enable transformative planetary science that would not only place vital constraints on the thermochemical evolution of Mercury, but would provide critical insight into the building blocks that formed the terrestrial worlds in this and other star systems. **We propose that the continued exploration of Mercury should be conceived as a multi-mission, multi-generational effort, guided by the crucial input provided by the Mercury science community, and that the landed exploration of Mercury be a high scientific priority in the coming decade.**

## 4. References

Albarède, F. (2009) Volatile accretion history of the terrestrial planets and dynamic implications. *Nature*, 461, 1,227–1,233, doi:10.1038/nature08477.
Anderson, B. J., et al. (2014) Steady-state field-aligned currents at Mercury. *Geophysical Research Letters*, 41, 7,444–7,452, doi:10.1002/2014GL061677.
Ash, M. E., Shapiro, I. I., and Smith, W. B. (1971) The system of planetary masses. *Science*, 174, 551–556, doi:10.1126/science.174.4009.551.
Banerdt, W. B., et al. (2012) InSight: An integrated exploration of the interior of Mars. *43rd Lunar and Planetary Science Conference*, abstract 2838.
Banks, M. E., et al. (2015) Duration of activity on lobate-scarp thrust faults on Mercury. *Journal of Geophysical Research Planets*, 120, 1,751–1,762, doi:10.1002/2015JE004828.
Benkhoff, J., et al. (2010) BepiColombo – Comprehensive exploration of Mercury: Mission overview and science goals. *Planetary and Space Science*, 58, 2–20, doi:10.1016/j.pss.2009.09.020.
Blewett, D. T., et al. (2011) Hollows on Mercury: Evidence for geologically recent volatile-related activity. *Science* 333, 1,856–1,859, doi:10.1126/science.1211681.
Blewett, D. T., et al. (2016) Analysis of MESSENGER high-resolution images of Mercury's hollows and implications for hollow formation. *Journal of Geophysical Research Planets*, 121, 1,798–1,813, doi:10.1002/2016JE005070.
Butler, B. J., Muhleman, D. O., and Slade, M. A. (1993) Mercury: Full-disk radar images and the detection and stability of ice at the north pole. *Journal of Geophysical Research*, 98, 15,003–15,023, doi:10.1029/93JE01581.
Byrne, P. K., et al. (2014) Mercury's global contraction much greater than earlier estimates. *Nature Geoscience*, 7, 301–307, doi:10.1038/ngeo2097.
Byrne, P. K., et al. (2016) Widespread effusive volcanism on Mercury likely ended by about 3.5 Ga. *Geophysical Research Letters*, 43, 7,408–7,416, doi:10.1002/2016GL069412.
Chabot, N. L., et al. (2016) Imaging Mercury's polar deposits during MESSENGER's low-altitude campaign. *Geophysical Research Letters*, 43, 9,461–9,468, doi:10.1002/2016GL070403.
Chabot, N. L., Shread, E. E., and Harmon, J. K. (2018) Investigating Mercury's south polar deposits: Arecibo radar observations and high-resolution determination of illumination conditions. *Journal of Geophysical Research Planets*, 123, 666–681, doi:10.1002/2017JE005500.
Dehant, V., et al. (2011) Revealing Mars' deep interior: Future geodesy missions using radio links between landers, orbiters, and the Earth. *Planetary and Space Science*, 59, 1,069–1,081, doi:10.1016/j.pss.2010.03.014.
Denevi, B. W., et al. (2013) The distribution and origin of smooth plains on Mercury. *Journal of Geophysical Research Planets*, 118, 891–907, doi:10.1002/jgre.20075.
Deutsch, A. N., et al. (2016) Comparison of areas in shadow from imaging and altimetry in the north polar region of Mercury and implications for polar ice deposits. *Icarus*, 280, 158–171, doi:10.1016/j.icarus.2016.06.015.
Deutsch, A. N., et al. (2020) Science Opportunities offered by Mercury's Ice-bearing Polar Deposits. *2023–2032 Decadal Survey White Paper*.
Domingue, D. L., et al. (2014) Mercury's weather-beaten surface: Understanding Mercury in the context of lunar and asteroidal space weathering studies. *Space Science Reviews*, 181, 121–214, doi:10.1007/S11214-014-0039-5.
Drinkwater, M., Floberghagen, R., Haagmans, R., Muzi, D., and Popescu, A. (2003) GOCE: ESA's first Earth Explorer core mission. *Space Science Reviews*, 108, 419–432, doi:10.1007/978-94-017-1333-7_36.
Ebel, D. S., and Stewart, S. T. (2018) The elusive origin of Mercury. In *Mercury: The View after MESSENGER*, ed. Solomon, S. C., Nittler, L. R., and Anderson, B. J. Cambridge Planetary Science, pp. 496–514.
Eng, D. A. (2018) Mercury lander mission concept study summary. *Mercury: Current and Future Science*, abstract 6070.
Ernst, C. M., et al. (2020) Mercury Lander Planetary Mission Concept Study. Prepared for the 2023–2032 Decadal Survey.

Ferrari, S., et al. (2014) In-situ high-temperature emissivity spectra and thermal expansion of C2/c pyroxenes: Implications for the surface of Mercury. *American Mineralogist*, 99, 786–792, doi:10.2138/am.2014.4698.

Genova, A., et al. (2019) Geodetic evidence that Mercury has a solid inner core. *Geophysical Research Letters*, 46, 7, 3625-3633, doi:10.1029/2018gl081135.

Gladman, B., and Coffey, J. (2009) Mercurian impact ejecta: Meteorites and mantle. *Meteoritics and Planetary Science*, 44, 285–291, doi:10.1111/j.1945-5100.2009.tb00734.x.

Griggs, C. E., et al. (2015) Tunable superconducting gravity gradiometer for Mars climate, atmosphere, and gravity field investigation. *46th Lunar and Planetary Science Conference*, abstract 1735.

Harmon, J. K., and Slade, M. A. (1992) Radar mapping of Mercury: Full-disk images and polar anomalies. *Science*, 258, 640–643, doi:10.1126/science.258.5082.640.

Harmon, J. K., Slade, M. A., and Rice, M. S. (2011) Radar imagery of Mercury's putative polar ice: 1999–2005 Arecibo results. *Icarus*, 211, 37–50, doi:10.1016/j.icarus.2010.08.007.

Hauck, S. A., II, Eng, D. A., and Tahu, G. J. (2010) Mercury Lander Mission Concept Study. Washington, DC: National Aeronautics and Space Administration.

Helbert, J., et al. (2013) Olivine thermal emissivity under extreme temperature ranges: Implication for Mercury surface. *Earth and Planetary Science Letters*, 371–372, 252–257, doi:10.1016/j.epsl.2013.03.038.

Hood, L. L. (2016) Magnetic anomalies concentrated near and within Mercury's impact basins: Early mapping and interpretation. *Journal of Geophysical Research Planets*, 121, 1,016–1,025, doi:10.1002/2016JE005048.

Iess, L., Asmar, S. and Tortora, P. (2009) MORE: An advanced tracking experiment for the exploration of Mercury with the mission BepiColombo. *Acta Astronautica*, 65, 666–675, doi:10.1016/j.actaastro.2009.01.049.

James, P. B., Zuber, M. T., Phillips, R. J., and Solomon, S. C. (2015) Support of long-wavelength topography on Mercury inferred from MESSENGER measurements of gravity and topography. *Journal of Geophysical Research: Planets*, 120, 287–310, doi:10.1002/2014JE004713.

Johnson, C. L., et al. (2015) Low-altitude magnetic field measurements by MESSENGER reveal Mercury's ancient crustal field. *Science*, 348, 892–895, doi:10.1126/science.aaa8720.

Johnson, C. L., et al. (2016) MESSENGER observations of induced magnetic fields in Mercury's core. *Geophysical Research Letters*, 43, 2,436–2,444, doi:10.1002/2015GL067370.

Lawrence, D. J., et al. (2013) Evidence for water ice near Mercury's north pole from MESSENGER Neutron Spectrometer measurements. *Science*, 339, 292–296, doi:10.1126/science.1229953.

Marchi, S., et al. (2013) Global resurfacing of Mercury 4.0–4.1 billion years ago by heavy bombardment and volcanism. *Nature*, 499, 59–61, doi:10.1038/nature12280.

Margot, J.-L., Hauck, S. A., II, Mazarico, E., Padovan, S., and Peale, S. J. (2018) Mercury's internal structure. In *Mercury: The View after MESSENGER*, ed. Solomon, S. C., Nittler, L. R., and Anderson, B. J. Cambridge Planetary Science, pp. 85–113.

Martinez, R., et al. (2017) Sputtering of sodium and potassium from nepheline: Secondary ion yields and velocity spectra. *Nuclear Instruments and Methods in Physics Research Section B: Beam Interactions with Materials and Atoms*, 406, 523–528, doi:10.1016/j.nimb.2017.01.042.

Mazarico, E., et al. (2014) The gravity field, orientation, and ephemeris of Mercury from MESSENGER observations after three years in orbit. *Journal of Geophysical Research Planets*, 119, 2,417–2,436, doi:10.1002/2014JE004675.

McCubbin, F. M., et al. (2017) A low O/Si ratio on the surface of Mercury: Evidence for silicon smelting? *Journal of Geophysical Research: Planets*, 122, 2,053–2,076, doi:10.1002/2017JE005367.

Merkel, A. W., et al. (2018) Evidence connecting Mercury's magnesium exosphere to its magnesium-rich surface terrane. *Geophysical Research Letters*, doi:10.1029/2018GL078407.

National Research Council (2011) *Vision and Voyages for Planetary Science in the Decade 2013–2022*. Washington, DC: The National Academies Press, doi:10.17226/13117.

Neumann, G. A., et al. (2013) Bright and dark polar deposits on Mercury: Evidence for surface volatiles. *Science*, 339, 296–300, doi:10.1126/science.1229764.

Nittler, L. R., et al. (2011) The major-element composition of Mercury's surface from MESSENGER X-ray spectrometry. *Science*, 333, 1,847–1,850, doi:10.1126/science.1211567.

Nittler, L. R., Chabot, N. L., Grove, T. L., and Peplowski, P. N. (2018) The chemical composition of Mercury. In *Mercury: The View after MESSENGER*, ed. Solomon, S. C., Nittler, L. R., and Anderson, B. J. Cambridge Planetary Science, pp. 30–51.

Padovan, S., Margot, J.-L., Hauck, S. A., II, Moore, W. B., and Solomon, S. C. (2014) The tides of Mercury and possible implications for its interior structure. *Journal of Geophysical Research Planets*, 119, 850–866, doi:10.1002/2013JE004459.

Paige, D. A., et al. (2013) Thermal stability of volatiles in the north polar region of Mercury. *Science*, 339, 300–303, doi:10.1126/science.1231106.

Peplowski, P. N., et al. (2011) Radioactive elements on Mercury's surface from MESSENGER: Implications for the planet's formation and evolution. *Science*, 333, 1,850–1,852, doi:10.1126/science.1211576.

Peplowski, P. N., et al. (2014) Enhanced sodium abundance in Mercury's north polar region revealed by the MESSENGER Gamma-Ray Spectrometer. *Icarus*, 228, 86–95, doi:10.1016/j.icarus.2013.09.007.

Peplowski, P. N., et al. (2015a) Geochemical terranes of Mercury's northern hemisphere as revealed by MESSENGER neutron measurements. *Icarus*, 253, 346–363, doi:10.1016/j.icarus.2015.02.002.

Peplowski, P. N., et al. (2015b) Constraints on the abundance of carbon in near-surface materials on Mercury: Results from the MESSENGER Gamma-Ray Spectrometer. *Planetary and Space Science*, 108, 98-107, doi:10.1016/j.pss.2015.01.008.

Phillips, R. J., et al. (2018) Mercury's crust and lithosphere: Structure and mechanics. In *Mercury: The View after MESSENGER*, ed. Solomon, S. C., Nittler, L. R., and Anderson, B. J. Cambridge Planetary Science, pp. 52–84.

Santerne, A., et al. (2018) An Earth-sized exoplanet with a Mercury-like composition. *Nature Astronomy*, 2, 393–400, doi:10.1038/s41550-018-0420-5.

Slade, M. A., Butler, B. J., and Muhleman, D. O. (1992) Mercury radar imaging: Evidence for polar ice. *Science*, 258, 635–640, doi:10.1126/science.258.5082.635.

Smith, D. E., et al. (2012) Gravity field and internal structure of Mercury from MESSENGER. *Science*, 336, 214–217, doi:10.1126/science.1218809.

Solomon, S. C., et al. (2008) Return to Mercury: A global perspective on MESSENGER's first Mercury flyby. *Science*, 321, 59–62, doi:10.1126/science.1159706.

Stanley, S., Bloxham, J., Hutchinson, W. E., and Zuber, M. T. (2005) Thin shell dynamo models consistent with Mercury's weak surface magnetic field. *Earth and Planetary Science Letters*, 234, 27–38, doi:10.1016/j.epsl.2005.02.040.

Strauss, B. E., Feinberg, J. M., and Johnson, C. L. (2016) Magnetic mineralogy of the Mercurian lithosphere. *Journal of Geophysical Research: Planets*, 121, 2,225–2,238, doi:10.1002/2016JE005054.

Vander Kaaden, K. E., et al. (2017) Geochemistry, mineralogy, and petrology of boninitic and komatiitic rocks on the mercurian surface: Insights into the mercurian mantle. *Icarus*, 285, 155–168, doi:10.1016/j.icarus.2016.11.041.

Vander Kaaden, K. E., et al. (2020) Mercury Sample Return to Revolutionize our Understanding of the Solar System. *2023–2032 Decadal Survey White Paper*.

Watters, T. R., et al. (2015) Distribution of large-scale contractional tectonic landforms on Mercury: Implications for the origin of global stresses. *Geophysical Research Letters*, 42, 3,755–3,763, doi:10.1002/2015gl063570.

Watters, T. R., et al. (2016) Recent tectonic activity on Mercury revealed by small thrust fault scarps. *Nature Geoscience*, 9, 743–747, doi:10.1038/ngeo2814.

Weider, S. Z., et al. (2015) Evidence for geochemical terranes on Mercury: Global mapping of major elements with MESSENGER's X-Ray Spectrometer. *Earth and Planetary Science Letters*, 416, 109–120, doi:10.1016/j.epsl.2015.01.023.

Williams, J.-P., Ruiz, J., Rosenburg, M. A., Aharonson, O., and Phillips, R. J. (2011) Insolation driven variations of Mercury's lithospheric strength. *Journal of Geophysical Research: Planets*, 116, E01008, doi:10.1029/2010JE003655.

Zhang, N., and Pommier, A. (2017) Electrical investigation of metal-olivine systems and application to the deep interior of Mercury. *Journal of Geophysical Research: Planets*, 122, 2,702–2,718, doi:10.1002/2017JE005390.

Zolotov, M. Y., et al. (2011) Implications of the MESSENGER discovery of high sulfur abundance on the surface of Mercury. *EOS* (Transactions, American Geophysical Union) American Geophysical Union, San Francisco, CA, pp. abstract # P41A-1584.

Zuber, M. T., et al. (2013) Gravity field of the Moon from the Gravity Recovery and Interior Laboratory (GRAIL) mission. *Science*, 339, 668–671, doi:10.1126/science.1231507.